\documentclass[preprint,aps,a4paper]{revtex4}
\usepackage{graphicx}

\begin{document}

\title{Does host contact structure reduce pathogen diversity?}
\author{A. Nunes$^{a*}$,  M. M. Telo da Gama$^{a}$ and  M. G. M. Gomes$^b$}
\address{$^a$Centro de F{\'\i}sica Te{\'o}rica e Computacional and
Departamento de F{\'\i}sica, Faculdade de Ci{\^e}ncias da Universidade de 
Lisboa, P-1649-003 Lisboa Codex, Portugal\\
$^b$Instituto Gulbenkian de Ci\^encia, Apartado 14, 2781-901 Oeiras, Portugal
\\$^*$Corresponding author: anunes@ptmat.fc.ul.pt}

\begin{abstract}

We investigate the dynamics of a simple epidemiological model 
for the invasion by a pathogen strain of a population where another  
strain circulates. We assume that reinfection by the same strain is 
possible but occurs at a reduced rate due to acquired immunity. The rate 
of reinfection by a distinct strain is also reduced due to cross-immunity.  
Individual based simulations of this model on a `small-world' network show 
that the host contact network structure significantly affects the outcome 
of such an invasion, and as a consequence will affect the patterns of 
pathogen evolution. In particular,
host populations interacting through a 'small-world' network of contacts 
support lower prevalence of
infection than well-mixed populations, and the region in parameter space
for which an invading strain can become endemic and coexist with 
the circulating strain is smaller,
reducing the potential to accommodate pathogen diversity. We
discuss the underlying mechanisms for the reported effects, and we propose
an effective mean-field model to account for the contact structure of the
host population in 'small-world' networks.
\end{abstract}

\maketitle
  

\vskip2pc

\section{Introduction}

Pathogens that diversify and evolve over relatively short time scales are
associated with specific features of infectious disease epidemiology.
Epidemics of acute respiratory infection occuring each winter in temperate
climates are caused in general by influenza (Hay 2001) and respiratory
syncytial viruses (Cane 2001). Rotaviruses (Iturriza-G\'omara 2004) are the
single most common cause of acute infantile gastroenteritis throughout the
world. These viruses undergo antigenic drift and both influenza A
and rotaviruses undergo major antigenic shifts when a new
virus is introduced into the human population through zoonotic transmission.
The dynamics of the epidemic is closely related to the antigenic
structure and evolution of the viral population. The spreading of multiple
strains may be investigated by means of mathematical models, and four
studies have recently uncovered the role of a `reinfection
threshold' (Gomes {\it et al} 2005) in regulating pathogen diversity (Gomes 
{\it et al} 2002; Boni {\it et al} 2004; Abu-Raddad \& Ferguson 2004;
G\"okaydin {\it et al} 2005). In spite of the significant differences
in model formulations, assumptions, and propositions, there is a striking
convergence at the level of the underlying mechanisms and results: levels 
of infection increase abruptly as a certain threshold is crossed, 
increasing the potential for pathogen diversity. This may 
be achieved by increasing the
transmissibility (modelled by $R_0$ in this work), by decreasing the
strength of the immunity (modelled by increasing $\sigma$ in this work) or by
increasing the rate at which novel pathogen variants are generated (not
considered in this work). 

The concept of reinfection threshold was introduced in
the context of 
a mean field model for vaccination impact  
(Gomes {\it et al} 2004) and has been the subject of 
some debate 
(Breban and Blower 2005), on the grounds that it corresponds to a change in
the order of magnitude of the equilibrium density of infectives that takes 
place
over a certain range of parameter values, rather than at
a well defined threshold.
In contrast with other phenomena in epidemic models where the term 
threshold is widely used, the reinfection threshold does not 
correspond to a bifurcation, or a phase transition, 
occurring at a well defined critical value. In Section 2, we give a precise
definition of the reinfection threshold concept, relating it to a bifurcation
that takes place in the limit when the demographic renewal rate tends to zero.
Thus, the reinfection threshold is a smoothed transition 
that occurs  in systems with
low birth and death rates, where a pronounced change
in the systems' steady state values takes place over a narrow 
range in
parameter space, akin to a smoothed  phase transition in finite physical 
systems around the value of the
critical control parameter where
a sharp
transition occurs in an
infinite system. This approach reconciles the two 
opposing views
in the literature, and we shall use the term reinfection threshold throughout
the paper in this sense.

For the invasion by a pathogen strain of a homogeneously mixed population 
where another strain circulates, it was recently found that strain 
replacement is favoured when the transmissibility is below the reinfection 
threshold, while coexistence is favoured above (G\"okaydin {\it et al} 2005). 
Here we consider a model of contact structure 
in the host population to describe the underlying network for
disease transmission, and investigate how this impacts on the outcome of an 
invasion by a new strain, that starts by infecting a very small fraction 
of the population.

Regular lattices are simple models that account for the geographical 
distribution of the host
population through local correlations but are not good descriptions of
social contact networks as they neglect mobility patterns. By contrast,
random graphs are simple models of complex networks that account for the
mobility of the host population, through the random links or connections,
but by neglecting local correlations 
render these models
poor descriptions of real social networks. 
A more realistic model of social networks was proposed recently by Watts 
and Strogatz. In
1998 they introduced small-world networks with topologies that
interpolate between lattices and random graphs (Watts \& Strogatz 1998). In
these networks a fraction of the lattice links is randomized by connecting
nodes, with probability $p$, with arbitrary nodes on the lattice; the number
of links is preserved by removing lattice links when random ones are
established. This interpolation is non linear: for a range of $p$ the
network exhibits small-world behaviour, where a local neighbourhood (as in
lattices) coexists with a short average path length (as in random graphs).
Analysis of real networks (Dorogotsev \& Mendes 2002) reveals the existence
of small-worlds in many interaction networks, including networks of social
contacts. In a recent work the contact patterns for contagion in urban
environments have been shown to share the essential characteristics of the
Watts and Strogatz small-world model (Eubank {\it et al} 2004).

In this work, we
implement the infection dynamics on a small-world network 
and find that the contact structure of the host population,
parametrized by $p$,
plays a crucial role in the final outcome of the invasion by a new strain, 
as
departures from the homogeneously mixed regime  favour strain
replacement and extinction versus coexistence (Section 3). Within the scope 
of an effective mean-field
model (Section 4), this effect is attributed to a reduction of the effective
transmissibility due to screening of infectives, resulting in a
translocation of the reinfection
threshold to higher transmissibility and thus decreasing the range where
strain coexistence is the preferred behaviour.

This conclusion seems to contradict the established idea that localized
interactions promote diversity, in the sense that the coexistence of
several competing strains is favoured with respect to homogeneously
mixed models. For the problem that we address here, this idea
was confirmed in a recent work (Buckee {\it et al}
2004) where strain competition in small-world host networks has been
considered. In Buckee {\it et al}, individual based stochastic simulations 
for a model with
short term immunity, and thus high rates of supply of naive susceptibles, 
have shown that a localized host contact structure favours pathogen diversity,
by 
reducing the spread of acquired immunity throughout the population.
Our results indicate that when the rate of supply of susceptibles
is low enough, the effect of localized interactions
reported by Buckee {\it et al}  
is superseded by that of the reduction of the effective transmissibility,
that places all but the most infectious strains below the reinfection 
threshold. Thus, in the case of long lasting partial immunity, 
a predominantely local contact structure will contribute to reduce
the number of coexisting competing strains in the host population.

\section{Analysis of the mean field model}

We consider an antigenically diverse pathogen population with a two-level
hierarchical structure: the population accommodates a number of strains (two
in this work); each strain consists of a number of
variants (here conceptualised as many). Variants in the same strain differ
by some average within-strain antigenic distance, $\sigma $, and different
strains are separated by some antigenic distance, 
$\sigma
_{\times}$. Antigenic distance is negatively related to cross-immunity and, in
some appropriate normalised measure, we may take $0\leq \sigma <\sigma 
_{\times}\leq 1
$. These assumptions are consistent with the idea that rapidly evolving RNA
viruses experience selection as groups of antigenically close strains
(Levin {\it et al} 2004). Studies of influenza A viral sequence evolution 
(Plotkin {\it et al} 2002; Smith {\it et al} 2004) 
provide empirical support of this modelling approach.

We consider a community of $N$ (fixed) individuals,  with
susceptibles and infectives in circulation. We assume that hosts are born
fully susceptible and acquire a certain degree of immunity as they are
subsequently infected. Individuals are characterized by their present
infection status (healthy or infected by strain $k$) and their history of
past infections (previously infected by a set of strains). In this model 
the dynamics of a single viral strain, representing a group of 
cocirculating antigenically close strains, is described by a 
Susceptible-Infected-Recovered (SIR) model 
with partial immunity against reinfection.
Abu-Raddad \& Ferguson (2004) have proposed a different modelling approach 
based on the reduction of a multi-strain system to an `equivalent' 
single strain SIR model, with parameters calculated to reproduce the 
values of the total disease prevalence and incidence of the original 
multi-strain model. In our model, the overall performance
of a strain depends on a single parameter, the degree of immunity,
measured by  $\sigma $, for
subsequent infections,  and the role and meaning of the reinfection threshold 
(see below for a definition)
become particularly clear.

The densities (or fractions of the host population) are denoted by $s_{j}$
for susceptibles and $i_{j}^{k}$ for infectives. The superscript $k$ of $i$ denotes the viral
strain that currently infects that fraction of the population. The subscripts $j$
describe the history of past infections (in the case of infectives, 
if different from the current infecting strain) and, for the model with two viral strains,
run over the subsets of $\{1,2\}$, that we will represent by $j=0,1,2,12$,
for the susceptibles. For the infectives, $j$ takes the values $0$, $1$ or $2$. 

The total fraction of the population infected by a given viral strain $k$ is 
$i^{k}=i_{0}^{k}+i_{j}^{k}$, $k,j =1,2$, $k\neq j$, 
and infection of fully susceptible individuals occurs at a rate per 
capita (force of infection) given by $\beta_{k}i^{k}$.
The parameter $\beta_{k}$ is the transmissibility for strain $k$.
Previous infections by the same strain reduce the transmissibility by a
factor $\sigma$ while previous infections by a different strain reduce the
transmissibility by a factor $\sigma_{\times}$. Constant and equal birth/death
rates, $\mu$, and constant recovery rate $\gamma$, are also assumed.
 
The mean-field equations for the dynamics of the various densities of
susceptibles and infectives are

\begin{eqnarray}
\frac{d s_0 }{dt} & = & \mu (1 - s_0) - (\beta_1 i^1 + \beta_2 i^2) s_0 
\nonumber \\
\frac{d s_j }{dt} & = & - \mu s_j + \gamma i_0^j - (\sigma_{\times} \beta_k i^k +
\sigma \beta_j i^j) s_j \,\,\,\,\,\,\,\,\, j,k=1,2;\,\,  j\neq k 
\nonumber \\
\frac{d i_0^k }{dt} & = & -(\gamma +\mu) i_0^k + \beta_k i^k ( s_0 
+ \sigma s_k ) \,\,\,\,\,\,\,\,\,\,\,\,\, k=1,2
\nonumber \\
\frac{d i_j^k }{dt} & = & -(\gamma +\mu) i_j^k + \beta_k i^k ( \sigma_{\times} s_j
+ \sigma s_{12} )
  \,\,\,\,\,\,\,\,\,\, j,k=1,2;\,\,  j\neq k
\label{mf}
\end{eqnarray}
where the density of susceptibles that were previously infected with both
viral strains is $s_{12}=1-s_0-s_1-s_2-i_1-i_2$. The parameter $\mu$ is
fixed at $\mu=1/80 yr^{-1}$ corresponding to a life expectancy of 80 years
and the rate of recovery from infection is also fixed at $\gamma= 52 yr^{-1}$
representing an average infectious period of approximately one week.

It is well known (Anderson \& May 1991) that a given single strain, $k$, 
persists at endemic equilibrium if $\beta_{k}>\gamma
+\mu$  or, equivalently, $R_{k}>1$ where $R_{k}$ is the basic reproduction
number of strain $k$,  
\begin{equation}
R_{k}=\frac{\beta _{k}}{\gamma +\mu }.
\end{equation}
The type of endemic equilibrium for the system described by (\ref{mf}) 
depends on the values of the 
transmissibilities, $\beta _{k}$ (or $R_{k}$). To establish whether 
two strains cocirculate at equilibrium, we evaluate the eigenvalues of the 
Jacobian at the single strain endemic equilibrium. 
If at least one of the eigenvalues is positive, the single strain 
equilibrium is unstable, and either coexistence or the other strain alone 
is the stable state. Applying this to strain 1, we find that this strain is 
expected 
to circulate alone when the pair $(R_{1},R_{2})$ is in the region marked 1 
in Figure 1(a,b).
Likewise, strain 2 is expected to circulate alone 
in region 2. The regions shaded in grey are
then identified as the set of parameter values for which
coexistence of both strains is stable, or coexistence regions.
For comparison, we include the dashed lines that delimit
the wider coexistence regions in the absence of reinfection by the 
same strain ($\sigma =0$).
The two panels correspond to systems with $\sigma = 0.25$, and 
different values of $\sigma_{\times}$: 
$\sigma_{\times}=0.27$  (or $\delta \sigma := \sigma _{\times}- \sigma =0.02$) 
in Figure 1(a); and $\sigma_{\times}=0.45$ (or $\delta \sigma =0.2$) in 
Figure 1(b). Comparing the two 
panels we make the expected observation that the stability of the coexisting 
solution is enhanced when the two strains are distantly related. We note 
the existence of 
two small `shoulders' or `kinks', marked A and B, on the boundaries
of the coexistence regions. By numerical inspection, the kinks are 
found to be close to

\begin{eqnarray}
A: & & R_1=\frac{1}{\sigma }, R_2=\frac{1}{\sigma_{\times}}, 
\nonumber \\
B: & & R_1=\frac{1}{\sigma_{\times}},  R_2=\frac{1}{\sigma }.
\end{eqnarray}

This makes intuitive sense. A simple SIR model with partial immunity 
(where reinfection occurs at a rate
reduced by $\sigma$) and with a low birth and death rate $\mu$
(as considered here) predicts a sharp increase in the density of infected 
individuals at $R_0$ close to $1/\sigma $ (Gomes 
{\it et al} 2004, 2005). This region of abrupt change
 is related to a
bifurcation at $R_0 = 1/\sigma $ that occurs when $\mu = 0$. We shall use 
the
term reinfection threshold for the smoothed transition that takes place at
small $\mu$, as well as for the approximate locus $R_0 = 1/\sigma $ of 
this
crossover. Focussing on strain 1, we expect a steep increase in
the infections by strain 1 as $R_1$ increases beyond $R_1=1/\sigma$, 
enhancing the competitive advantage of strain 1 relative to strain 2 and
reducing the coexistence region. Two limiting cases deserve special
reference: when $\sigma$ is as large as $\sigma_{\times}$,  $A=B$ and the
coexistence region collapses to the diagonal; and when $\sigma\rightarrow 0$ 
as in previously studied models
(May \& Anderson 1983; Bremerman \& Thiemme 1989) the coexistence region
is maximal.

Hereafter we will restrict ourselves to a symmetrical model,
and will denote $R_1=R_2$ by the usual symbol for the 
basic 
reproduction
number, $R_0$. Figure 1(c,d) shows the stability of the coexistence
equilibrium as a function of $R_0$, measured by the real parts of the
seven eigenvalues of the linear approximation of system (\ref{mf})
at equilibrium for $\sigma =0.25$ and $\delta \sigma =0.02$
(Figure 1(c)), $\delta \sigma =0.2$ (Figure 1(d)). All the eigenvalues
have negative real parts, and all but the smallest of them change by 
several orders
of magnitude as $R_0$ crosses the reinfection threshold region
around $R_0 = 1/ \sigma$. The plot shows the modulus of the real
parts of the seven eigenvalues, in logarithmic scale,   
versus $R_0$, and the vertical line through $R_0 = 1/ \sigma$ marks the 
position of the reinfection threshold. Also shown by the dashed line
is the total density of infectives at the equilibrium, which
also increases by two orders of magnitude as $R_0$ crosses the 
reinfection threshold region.

In both cases we observe that as the transmissibility crosses the
reinfection threshold around $R_0=1/\sigma$ the overall stability of the
coexistence equilibrium increases significantly, as 
the largest eigenvalues decrease steeply. This effect is achieved by 
increasing either $R_0$ or $\sigma$, and the system behaviour can be 
characterised in 
terms of $R_0$ or of $\sigma$. Throughout the rest of
this work we fix $R_0$ and vary $\sigma$.

The behaviour of the solutions of (\ref{mf})
 for $\delta \sigma = \sigma_{\times} - \sigma = 0.02$
and $\delta \sigma = 0.2$ is shown in Figures 2 and 3. We have taken $\sigma
= 0.2$, $\sigma = 0.25$ and $\sigma = 0.3$ in panels a), b) and c),
respectively, corresponding to below threshold, threshold and above
threshold behaviour for the fixed value of $R_0 = 3.9$. 
By threshold we mean, as before, the window of parameter values 
around $R_0 = 1/ \sigma$ that corresponds to the smoothed transition
that takes place for small birth and death rate $\mu $, where
the systems' coexistence equilibrium densities and stability properties
change very rapidly. 
This is illustrated
in the panels d), e) and f) of Figures 2 and 3, where the dashed line is
at $R_0 =3.9$ and the dotted line indicates the position of the reinfection
threshold $R_0 = 1/ \sigma$, for $\sigma =0.2$ (Figure 2(d) and 3(d)), 
$\sigma =0.25$ (Figure 2(e) and 3(e)) and $\sigma =0.3$ (Figure 2(f) and 3(f)).
The full line depicts the total equilibrium density
of infectives with either one of the two strains
as a function of $R_0$ for the (unique) stable steady state of
the system, corresponding to endemic equilibrium where both viral strains
coexist at the same steady state densities.

The numerical integration of equations (\ref{mf}) starts with a single 
circulating strain (strain 1), and when the single strain steady
state is reached at $t\sim 20$ years
a small fraction ($1.5 \times 10^{-4}$) of individuals infected with the invading 
strain
(strain 2) is introduced in the system. The two curves in panels a), b) and c)
of Figures 2 and 3 represent, in logarithmic scale, the infective densities 
for each of the strains as a function of time obtained from this numerical 
integration. The curve that corresponds to the density of infectives
carrying strain 2 starts at $t\sim 20$ years with density
$1.5 \times 10^{-4}$ in all the examples shown.

On relatively short time scales of a few decades, we find that the
reinfection threshold is the boundary between two different regimes. For
very similar viral strains, as depicted in Figure 2, the behaviour of the
model below threshold is strain replacement, and the behaviour on and
above threshold is strain coexistence. For distinct viral strains, as in
Figure 3, the outcome is always strain coexistence, but the
density oscillations are negligible above threshold, and very pronounced
below threshold. In order to extract epidemiologically significant 
predictions from the mean
field model we have to take into account that the population is discrete,
and that densities below a certain lower bound, that depends on the
population size, are effectively zero. If we take this effect into account
and set the densities to zero when they fall below $10^{-6}$, we obtain 
extinction of both strains in the case of Figure
3(a).  

For longer time scales, however, the model also predicts
coexistence for the case of  similar strains below threshold
(Figure 2(a)) since there 
are no attractors in the single strain invariant subspaces when the two strains
have the same transmissibility. All the solutions that were analysed
tend eventually to the coexistence equilibrium, but the transients for 
similar strains below threshold last for several hundred years, during
which the density of infectives carrying strain 1 reaches much smaller 
values than the cut-off of $10^{-6}$.

This behaviour and the role of the reinfection threshold can be interpreted
in terms of the eigenvalues of the coexistence endemic equilibrium
represented in Figure 1(c,d). As $R_0$ increases across the threshold,
the real parts of the eigenvalues decrease steeply, and
the imaginary parts vanish. As a result the
oscillatory component of the system plays an important role below the 
reinfection threshold and is absent above. This is the basic
ingredient of the different invasion dynamics depicted in Figures 2 and 3.

The difference in below threshold behaviour between
similar and dissimilar strains, strain replacement in the
former case and total extinction in the latter may be 
understood intuitively in terms of the reduced cross-immunity
of dissimilar strains. In this
case, the small fraction of infectives carrying
strain 2 finds a highly susceptible population and generates
a large epidemic outbreak, during which strain 1 goes extinct
because of the lack of suceptibles. Strain 2
quickly exhausts its pool of susceptibles through first infections,
and then dies out too as $\sigma R_0 <1$.

Then, for realistic
population sizes, the predictions of the mean-field model are the following.
For similar viruses, strain replacement (drift) occurs below threshold, and 
coexistence on and
above threshold. For dissimilar viruses, global extinction occurs below
threshold while coexistence occurs on and above threshold. 

It is known (Bayley 1975) that fluctuations due to stochastic 
effects in discrete populations will also
favour strain extinction and replacement with respect to deterministic model
predictions. A homogeneously mixed stochastic version of 
model (\ref{mf}) was considered by G\"okaydin {\it et al} 
(2005) for population sizes between $10^6$ and $10^8$, and, as expected, 
stochasticity was found to favour 
strain replacement and extinction both for similar and dissimilar 
strains. This effect may be, 
however, greatly enhanced in more realistic descriptions of the host 
population where the homogeneously mixed assumption is relaxed.

\section{Drift and shift in small-world networks}

In the presence of random long-range links such as those considered in
small-world networks the endemic and epidemic thresholds of 
Susceptible-Infective-Susceptible (SIS) and SIR
models may be mapped on to {\it mean-field} site and bond percolation
transitions (Grassberger 1983, Dammer \& Hinrichsen 2003, Hastings 2003).
In recent works, the network topology has been considered in the 
calculation of
endemic and epidemic thresholds of SIS and SIR models (May \& Lloyd 2001,
Moore \& Newmann 2000a, 2000b, Pastor-Santorras \& Vespigniani 2001a) and
the results revealed a strong dependence of the threshold values on the
network size and structure.
The contact network topology has also been shown to play an important 
role in the
stationary
properties of the endemic state (Pastor-Santorras \& Vespigniani 2001b), the
short term dynamics of epidemic bursts (Keeling 1999, Kleczkowski \&
Grenfell 1999, Rhodes \& Anderson 1996), the long term dynamics of childhood
diseases (Verdasca {\it et al} 2005) and in the estimation of disease
parameters from epidemiological data (Meyers {\it et al} 2005). 

The structure of the network of contacts of the host population
is also expected to play a role in the
invasion dynamics described in the previous section. This will affect the
pattern of strain replacement and as a consequence the evolution
of competing multi-strain 
pathogens. In
the following we analyse quantitatively the effects of a network of social
contacts with small-world topology on the simple two strain model
(\ref{mf}).

We implemented a discrete version of the model (\ref{mf}) on a cellular 
automaton
(square lattice) with $N= 800\times 800$ sites and small-world interaction 
rules (see the appendix for a description of the algorithm). We
started by simulating the behaviour of the system for a single strain model,
 whose mean field equations are simply
\begin{eqnarray}
\frac{d s_0 }{dt} & = & \mu (1 - s_0) - \beta i  s_0 
\nonumber \\
\frac{d i }{dt} & = & - (\gamma + \mu) i + \beta i  
(s_0 + \sigma (1- s_0 -i))
\label{mf1}
\end{eqnarray}
We found characteristic medium and long-term dynamics related, in a
quantitative fashion, to the structure of the network of contacts. In
particular, as $p$ decreases, the increase in spatial correlations (i)
decreases the effective transmissibility through the screening of
infectives and susceptibles, which in turn increases the value of the
transmissibility at the endemic and reinfection thresholds. In addition,
the spatial correlations (ii) enhance the stochastic fluctuations with
respect to the homogeneously mixed stochastic model. This effect is
particularly strong at low $p$, where the relative fluctuations are largest
and where as a consequence (iii) the dependence of the steady state
densities on the effective transmissibility predicted by the mean-field
equations breaks down. Analogous effects, including departures from the
mean-field behaviour at the endemic threshold (persistence transition) of a
SIR model with non-zero birth rate have been reported recently (Verdasca 
{\it et al} 2005).

In Figure 4 we plot the effective transmissibility, $\beta_{eff}$ (average
density of new infectives per time step divided by the value of $i(s_0
+\sigma(1-s_0-i))$ for the instantaneous densities), at fixed $R_0=3.9$ and 
$\sigma= 0.30$, well above the reinfection threshold of the mean-field model,
as a function of the small-world parameter, $p$. The variation of the
effective transmissibility with $p$ is due to the clustering of infectives
and susceptibles. This is a well known screening effect that results 
from the local structure (correlations), 
and has to be taken into account in fittings to effective
mean-field models. In this framework, the reduction of the 
effective transmissibility represents an increase of the value of $R_0$
at the reinfection threshold.

The clustering, and spatial correlations in general, have also drastic 
consequences on the amplitude and nature of the stochastic fluctuations.
Enhanced fluctuations will ultimately
lead to stochastic extinction as $p$ decreases, but before extinction
occurs, the fluctuations lead
to the appearance of a regime dominated by local structure and correlations,
where the mean-field relations among the equilibrium densities break down.
This is illustrated in Figure 5 where we plot the effective transmission
rate (dots) as a function of the equilibrium density of infectives (model 
parameters as
in Figure 4).
In the same figure we plot the equilibrium value of the
density of infectives (full line) predicted by the mean-field equations
(\ref{mf1})
for a value of the transmissibility equal to the effective
transmissibility obtained from the simulations. When $p>p_b$, away from the network
endemic threshold at $p=p_c$, the mean-field equilibrium density for the
effective transmissibility is in excellent agreement with the simulated
equilibrium density. In the region of small $p$ ($p_c < p < p_b$)), however, the simulated
equilibrium densities differ significantly from those calculated using the
effective mean-field theory. A departure from the mean-field dependence of
the effective transmissibility on the steady state density of infectives
implies that the description of the system is no longer possible in terms of
effective mean-field theories based on density dynamics, by contrast with the
second regime where the contact structure can be implicitely taken into
account by effective mean-field models (Section 4).

We then focussed on the analysis of the invasion dynamics. 
We performed
individual based simulations 
starting from an initial condition where the system is close to the
steady state for a single resident strain and introduced a small fraction of 
individuals 
($1.5 \times 10^{-4}$) infected by the invading strain. 
The algorithm is a natural extension of the single strain stochastic
algorithm for SIR dynamics with reinfection on a small-world
network of contacts, and a detailed description is given in the appendix. 
We found that as local
effects become important (low $p$), strain replacement (and also total 
extinction) are favoured with respect to the homogeneously mixed regime 
($p=1$). 
These results are summarized in Figure 6 where we plot $\sigma$ 
at crossover between the different regimes, coexistence, replacement
and extinction, versus the 
small-world parameter, $p$, for
systems where the competing viral strains are similar, $\delta \sigma=0.02$ 
(Figure 6(a))
and dissimilar, $\delta \sigma=0.2$ (Figure 6(b). The full line is the 
boundary between the strain coexistence and the strain replacement
regimes, and the dashed line is the boundary between 
strain replacement and total extinction.
The final outcome of an invasion is determined
by carrying out a series of simulations, for different values of $p$ and 
$\sigma$, and keping $R_0$ fixed at $3.9$. For each  
$(p,\sigma)$, twenty (forty in some cases) invasion simulations are performed
for a period of 13.7 years, with invasion at $t=4.1$years.
The fraction of simulations where both the circulating and the invading  
strain prevail
is plotted as a function of $\sigma$, at fixed $p$. This fraction changes
rapidly from one to zero as $\sigma $ decreases across a small interval, and
the boundary of the coexistence regime (the full line in Figure 6(a,b)
is determined as the point where
it takes the value $0.5$.
A similar analysis 
yields the boundary 
that separates the
replacement from the total extinction regime (the dashed line in Figure 
6(a,b). The results show that as $p$ decreases the value of $\sigma$ at these
two boundaries
increases exponentially from its value in the homogeneously 
mixed
system, $p=1$. As a consequence, the range of parameters in the coexistence
regime decreases drastically as $p$ decreases, supporting the claim that the
structure of the small-world network of contacts hinders, rather than favours, 
pathogen diversity.

This contrasts with the behaviour reported by Buckee {\it et al} 
(2004) for a model of strain evolution including short term host immunity 
and cross-immunity on a 
(static) small-world network of contacts, where coexistence of
competing strains was found to be favoured with respect to the homogeneously
mixed model. The different behaviour we report is due to the combined effect
of small-world network structure and a low rate of supply of naive susceptibles,
and may be understood in terms of the analysis of the single strain 
stochastic model discussed previously, together with the behaviour of the 
two-strain mean-field model which, for low $\mu $, predicts strain coexistence
as the outcome of an invasion only when the transmissibility is high enough. 

Recall that in the mean-field model of Section 2 (system (\ref{mf})) when a 
second strain is
introduced in a population with a resident virus, drift occurs below the 
reinfection threshold for antigenically similar
strains (small $\delta \sigma $). 
When the resident and the invading strains are antigenically distant 
(large $\delta \sigma $) the final outcome is coexistence at and 
above the reinfection threshold and global extinction below threshold.
The results of the individual based simulations show that the enhancement 
of the stochastic fluctuations due to spatial correlations favours
replacement where we would otherwise have coexistence, and global extinction 
where we would otherwise have replacement. 
In particular, replacement (both drift and shift, replacement
of dissimilar strains) becomes a typical 
outcome at the reinfection threshold, instead of coexistence as in the mean-field model. 
This may be viewed as a finite size stochastic effect similar to the 
effect reported in G\"okaydin {\it et al} 2005 for the homogeneously 
mixed stochastic model, or to what we have also found here for $p=1$, but 
much more pronounced. As we noted previously the amplitude of the 
stochastic fluctuations at
small values of $p$ is greatly enhanced with respect to the amplitude of the
stochastic fluctuations at $p=1$, due to coherent fluctuations of 
host population clusters. 

However, the main effect of spatial correlations that accounts for the results
of Figure 6 is the screening of infectives
that leads to the reduction of the effective transmissibility as shown
in Figure 4. 
As $p$ decreases, this effect  brings a system above the 
reinfection threshold closer to
or even below the reinfection threshold, also favouring replacement for
small $\delta \sigma$ and either replacement or extinction for larger 
$\delta \sigma$.
While the effects of the coherent fluctuations of clusters are important 
only in a small range of $p$ above the single strain endemic 
transition, screening of infectives occurs over the whole range of $p$ with 
the 
corresponding translocation of the reinfection threshold to larger values of 
$R_0$.
In particular, for a large range of values of $p$ we 
obtain drift/shift, instead of coexistence as predicted by the well mixed 
model for the same disease parameters. Many of these effects may be 
captured by effective mean-field models as described in the next section.

\section{An effective mean-field model}

In Section 3 we have seen that, for single strain dynamics, the mean
field equilibrium incidence given by (\ref{mf1}) for the effective 
transmissibility is
in excellent agreement with the results of the simulations for 
values of the small world parameter  $p$ in an interval $[p_{b}, 1]$,
while departure from the mean field relations occurs close to the 
endemic transition threshold at $p_c < p < p_b$ (see Figure 5).

In this section we show that, for the full system and $p$ in the range 
$[p_{b}, 1]$,
the simulated time series are well
approximated by the solutions of an effective mean-field model of
the form (\ref{mf}).
For the
construction of the effective model we assume that the transmissibility is
of the form $\beta _{eff}= \beta f(p)$, where $f(p)$ is a function, unknown
a priori, that represents the screening of infectives. The screening
function $f(p)$ is obtained from a plot $\beta _{eff} (p)$ as in Figure 4,
where the screeening effect is quantified for the single strain dynamics.
The relevant range of $p$ for this fit is $[p_{b}, 1]$, since the breakdown
of the mean field equilibrium relations implies that our ansatz for the form
of the infection rate is no longer valid.
Even in this range  it is not obvious that the
ansatz will yield an effective mean-field model capable of describing the full
dynamical behaviour of the simulations when a resident and an invading
strain interact. 

In Figure 7(a), black line, we show the results of stochastic 
invasion simulations as described in Section 3 and in the Appendix, for 
$p=0.5$ (other parameters as in
Figure 2.b)), together with 
the numerical integrations, grey line,
 of the coupled strain effective mean-field 
model, given by (\ref{mf}) with $\beta _k = \beta _{eff}$, for $k=1,2$.
Also shown in Figure 7(b) are, for comparison, the solutions of the standard 
mean-field model, where $f(p)=1$ (grey line), plotted together with the 
stochastic simulations of Figure 7(a) (black line). 
In all cases we use the same initial conditions and the same invasion
conditions.

While the standard mean-field model 
disagrees
quantitatively, and even qualitatively, with the results of the simulations,
the effective mean-field model describes accurately the dynamical 
behaviour
of the system, except for the stochastic fluctuations about the average
densities.

Systematic calculations show that the effective mean-field model performs
rather well against simulations, everywhere in the region where the
mean-field relations apply. In this regime, the effect of the contact
structure of the host population is accurately described through the
screening function $f(p)$, obtained from the steady state averages of single
strain simulations.

In the range $[p_c, p_b]$ of $p$ above the endemic threshold where we
have found significant departures from the mean-field description, one may
think that a more general ansatz for the form of the force of infection
could lead to a modified effective mean-field model, albeit a more
complicated one, that would fit the simulation results.
However, it is easy to check that the parametric plot of 
$\beta _{eff}(p)$ vs $i^*(p)$, where $i^*(p)$ is the steady state average
infective density, follows the mean field relation of the model with linear
force of infection whatever the functional form assumed for $\beta (i,p)$.
Indeed, for any function $\beta (i,p)$, and rate of infection of the form 
\begin{equation}
\beta (i,p)i(s + \sigma (1-s-i)),  \label{forcei}
\end{equation}
$\beta_{eff}(p)$ measured from the simulations is $\beta _{eff}(p) = \beta
(i^*,p)$, and the curve $(\beta _{eff}(p), i^*(p))$ is the same as the
standard mean field curve for $\beta (i^*)$.
This means that even in the scope of a more general model with a
nonlinear force of infection
the mean field curve for $\beta _{eff}(p)$ vs $i^*(p)$,
 the full line in Figure 5, does not fit
the simulations results for $p$ in the range $[p_c, p_b]$.

We conclude that the effect of the breakdown of the mean-field relations
reported in Section 3 is an indication of a new regime where spatial
correlations are too important for the contact rate between individuals of
different classes to be described by the product of the corresponding
densities. The construction of effective models in this regime requires the
use of pair approximations that take into account the spatial correlations
(de Aguiar {\it et al} 2003) and will be the subject of future work.

\section{Conclusions}

We performed stochastic simulations of individual based models on
small-world networks to represent host populations where one or two
viral strains, each representing a group of antigenically close variants, 
may be present. In the case of a single viral strain, the node dynamics 
corresponds to a SIR model with reinfection at a reduced rate due to acquired
 immunity. In the case of two competing
strains, the rate of infection by a distinct strain is also reduced due to 
cross 
immunity, which is always weaker than strain specific immunity.
Both the single and the double strain models include crucial ingredients 
required by realistic modeling of the host population,
namely stochasticity, a discrete finite population, and spatial structure
given by a plausible contact network. 

We analised both the reinfection dynamics of the single 
strain model and the invasion dynamics of the two strains model,
taking initial conditions that correspond to the presence of a small number
of individuals infected with a pathogen strain in a population where
another
pathogen strain circulates. 

For single-strain dynamics,
we found that the major effect of spatial correlations is a decrease in 
the
effective transmissibility through the screening of infectives and
susceptibles which in turn increases the value of the transmissibility at
the endemic and at the reinfection thresholds. In addition, spatial 
correlations
enhance the amplitude of the stochastic fluctuations with
respect to the homogeneously mixed stochastic model. This effect is
particularly strong at low $p$, where the relative fluctuations are largest
and where the mean-field dependence of the steady state densities on the
effective transmissibility breaks down.
Indeed, we have found a regime at small $p$, where spatial correlations 
dominate and the mean-field relations break down, as well as a
second, wider, regime for larger values of $p$ where effective mean-field
models are capable of describing the essential effects of the spatial
correlations through a reduced effective transmissibility.

For the two-strain model, we found that the host contact 
structure significantly affects the outcome of an attempted 
invasion by another strain of a population with an endemic resident strain, and 
by contrast with standard expectations we observed that spatial 
structuring reduces the potential for pathogen diversity. 
The simulation results show that the structure of the 
network of contacts favours strain replacement (drift/shift) or global 
extinction 
versus coexistence as the
outcome of an invasion, with respect to the mean-field and stochastic
homogeneously mixed populations. In particular, we find that as the
small-world parameter, $p$, decreases at fixed $R_0$, the 
value of the strain specific immunity parameter $\sigma $ 
above which coexistence is the  typical outcome
increases exponentially from its value for the homogeneously mixed
system, $p=1$, supporting the claim that local correlations strongly reduce 
pathogen diversity.

This conclusion is in apparent contradiction with the established idea 
that localized interactions promote diversity, as confirmed by stochastic 
simulations for a model with high rates of 
supply of naive susceptibles. In that model the effect of the host contact 
structure favours pathogen diversity, by suppressing the uniform spread of 
acquired immunity throughout the population. 
By contrast, in our model characterised by partial permanent 
immunity and low rates of supply of naive susceptibles (through births) 
the former effect is superseded by the screening effect. The latter
is also due to the clustering of the host population, and results in a 
reduction of the effective transmissibility, changing the reinfection 
threshold to higher levels of $R_0$, and therefore reducing the range where
 coexistence is the preferred behaviour. 

The results of our simulations strongly support the conclusion that in 
systems with a reinfection threshold, due to a low rate of supply of naive 
susceptibles, the major effect of 
the host population spatial structuring is the effective reduction of pathogen 
diversity.  

The immunity profile (Ferguson \& al. 2003) and the duration of infection
(Gog \& Grenfell 2002) have recently been shown to play an important
role in shaping the patterns of pathogen evolution in multi-strain
models with mutation generated antigenic diversity. Investigating
the additional effect of host population contact structure topology
in these models will be the subject of future work.
                           
\section*{Acknowledgements}

Financial support from the Portuguese Foundation for Science and Technology
(FCT) under contracts POCTI/ESP/44511/2002, POCTI/ISFL/2/618 and POCTI/MAT/47510/2002, and from the
 European Commission under grant MEXT-CT-2004-14338, is gratefully acknowledged.
The authors also acknowledge the contributions of J. P Torres and
M. Sim\~oes to test and to improve the code used in the simulations.

\newpage

\section{Appendix}
\subsection{description of the algorithm}

A community of $N$ (fixed) individuals comprises, at time $t$, 
susceptibles and infectives in circulation. Hosts are born fully 
susceptible and acquire a certain degree of immunity as they
are subsequently infected. We consider susceptibles and infectives that 
were previously infected by strain $i$, $S_i$ and $I^k_i$. The indices 
denote previous and current infections as in the text.

We consider a cellular automaton (CA) on a square lattice of size $N=L^{2}$
with periodic boundary conditions. The (random) variables at each site may
take one of eight values: $S_0$, $S_1$, $S_2$, $I_0^1$, $I_0^2$, $I_1^2$, 
$I_2^1$ or $S_{12}$. The lattice is full. We account for local
interactions/connections with $k$ neighbouring sites, with $k=12$, and
long-range interactions/connections, with a small-world probability, $p$.

The transmissibility $\beta$ is the sum of the local and long-range rates
of transmission. First infection, within-strain and between-strains 
reinfection, recovery, birth and death occur stochastically, with fixed 
rates ($\beta$, $\beta \sigma $, $\beta \sigma_{\times}$, $\gamma$, $\mu$, $\mu)$. 
The recovery time ($1 /\gamma$) sets the time scale. At each Monte 
Carlo (time) step, $N$ random
site updates are performed following a standard algorithm. The type of
event, long or short-range infection, within and between-strains 
reinfections by strains 1 and 2, recovery, birth and death, is chosen with 
the appropriate frequency ($\beta p$, $\beta (1-p)$, $\beta p \sigma $, 
$\beta (1-p)\sigma $, $\beta p \sigma_{\times}$, $\beta (1-p) \sigma_{\times}$, $\gamma$, 
$\mu$, $\mu$) and then proceeds as follows.

a. Long(short)-range first infection by 1 (2). A site is chosen at random; if the
site is occupied by an $I$ or an $S$ other than $S_0$ no action is taken. If
the site is occupied by an $S_0$, one of the other lattice sites (or of its $k$
neighbours for short-range infection) is chosen at random; $S_0$ is infected
by 1 (2) iff that site is an $I^1$ ($I^2$).

b. Long(short)-range between strain reinfection by 1 (2). A site is chosen at
random; if the site is occupied by an $I$ or an $S$ other than $S_2$ ($S_1$) 
no action is taken. If the site is occupied by an $S_2$ ($S_1$), one of the
other lattice sites (or of its $k$ neighbours for short-range infection) is
chosen at random; $S_2$ ($S_1$) is reinfected by 1 (2) iff that site is an $I^1$ ($I^2$).

c. Long(short)-range within strain reinfection by 1 (2). A site is chosen at
random; if the site is occupied by an $I$ or an $S$ other than $S_1$ ($S_2$)
or $S_{12}$ no action is taken. If the site is occupied by an $S_1$ ($S_2$)
or $S_{12}$ one of the other lattice sites (or its $k$ neighbours for
short-range infection) is chosen at random; $S_1$ ($S_2$) or $S_{12}$ is
reinfected by 1 (2) iff that site is an $I^1$ ($I^2$).

d. Recovery: a site is chosen at random; if the site is occupied by an $I$
recovery occurs.

e. Death and birth: a site is chosen at random and death occurs regardless;
the site is then (re-)occupied by an $S_0$. Death is linear in the densities
and birth is independent of the densities.

We used systems with $800 \times 800$ nodes.\\

\newpage

\centerline {FIGURES}

{\bf FIGURE 1} Stability analysis. (a,b) Conditions for strain coexistence
or competitive exclusion as described by the two strain mean-field model.
The panels correspond to models with $\sigma=0.25$ and different values of 
$\sigma_{\times}$: (a) $\sigma_{\times}=0.27$; (b) $\sigma_{\times}=0.45$. 
The regions shaded in grey are
the set of parameter values for which
coexistence of both strains is stable, or coexistence regions.
The dashed lines delimit
the wider coexistence regions in the absence of reinfection by the 
same strain ($\sigma =0$). In region 1 (resp. 2), the single
strain endemic equilibrium for strain 1 (resp. 2) is stable.
(c,d) 
Modulus of the real parts of 
the seven eigenvalues of the coexistence equilibrium along the diagonal of 
(a,b) respectively, in logarithmic scale. Also shown is the total density of infectives
at equilibrium (dashed line), and the position of the
reinfection threshold (dotted line). 

\bigskip

{\bf FIGURE 2} Mean-field new infectives densities for the two strain model
of equations (1-4) with $\delta \sigma = \sigma_{\times} - \sigma = 0.02$. We 
have
taken $\sigma = 0.2$, $\sigma = 0.25$ and $\sigma = 0.3$ in panels a), b)
and c), respectively, corresponding to below threshold, threshold and above
threshold behaviour for the value of $R_0 = 3.9$. This is illustrated in the
diagrams d), e) and f) where the dashed line is at $R_0 =3.9$ and the dotted
line indicates the position of the reinfection threshold $R_0 = 1/\sigma$.
The full line depicts the equilibrium total density of infectives, $i^*$, as a
function of $R_0$ for the stable steady state of the system, the endemic
equilibrium where both viral strains coexist. 

\bigskip

{\bf FIGURE 3} Mean-field new infectives densities for $\delta \sigma = 0.2$. 
As in Figure 2 we take $\sigma = 0.2$, $\sigma = 0.25$ and $\sigma = 0.3$
in panels a), b) and c), respectively, corresponding to below threshold,
threshold and above threshold behaviour for $R_0 = 3.9$, as illustrated in
the diagrams of panels d), e) and f).

\bigskip

{\bf FIGURE 4} Effective transmissibility vs the small-world parameter, $p$. 
The effective transmissibility, $\beta_{eff}$, is calculated as the
average density of new infectives per time step divided by 
$i(s_0+\sigma(1-s_0-i))$ where $s_0$ and $i$ are the instantaneous densities of
susceptibles and infectives. The drastic reduction in $\beta_{eff}$ is due
to the clustering of infectives and susceptibles as $p$ decreases and the
spatial correlations increase. The model is the single strain reinfection
model (\ref{mf1}) and the parameters are $R_0=3.9$ and $\sigma=0.3$. The number of
nodes is $800 \times 800$.

\bigskip



{\bf FIGURE 5} Effective transmissibility vs the equilibrium infective
density, $i^*(p)$ (dots), over the whole range of the small-world parameter, $p$
(dots). Notice the logarithmic scale in the $i^*(p)$ axis.
The full line is the $\beta(i^*)$ curve calculated from the
mean-field single strain reinfection model (\ref{mf1}). The parameters are as in
Figure 4. The results show the existence of two regimes: an effective
mean-field regime for $p > p_b$, where the relations between the equilibrium
densities are given by the mean-field equations for the screened value of
the transmissibility, and a fluctuation dominated regime for $p_c < p < p_b$,
where the mean-field relations break down.

\bigskip

{\bf FIGURE 6} Coexistence versus replacement (full line) and replacement
versus total extinction (dashed line) crossovers in $(p,\sigma)$ space. 
The reinfection parameter, 
$\sigma$,  is plotted as a function of the small-world
parameter, $p$, at the crossover separating the dynamical regimes of coexistence of two
competing viral strains (above the full line), of invasion prevalence
(between the dashed and the full lines) and of total extinction (below the
dashed line). See the text for details. Panel a) corresponds to similar
strains, $\delta\sigma=0.02$, and plot b) to dissimilar strains, 
$\delta\sigma=0.2$. $R_0 = 3.9$ and $N$ as in other simulations.

\bigskip

{\bf FIGURE 7} Mean field and individual based description of invasion
dynamics. The same initial conditions and the same invasion conditions
are used in all cases.
(a) Mean-field new infectives densities (grey line) for the model of
equations (\ref{mf}) with $\beta _k = \beta _{eff}$, $k=1,2$, 
fitted from the simulations of the single strain
model for $p=0.5$ (other parameters as in Figure 2(b)), and time series
(black line) of the stochastic model for $p=0.5$ 
(other parameters as in Figure 2(b)). (b) We also show, for
comparison, the solutions of the standard mean-field model (grey line), where 
$\beta _k = \beta $, $k=1,2$, are not corrected to take into account
the screening. The black line corresponds to the same data as in (a).

\newpage

\begin{figure}[tbp]
\begin{center}
\begin{tabular}{cc}
\includegraphics
{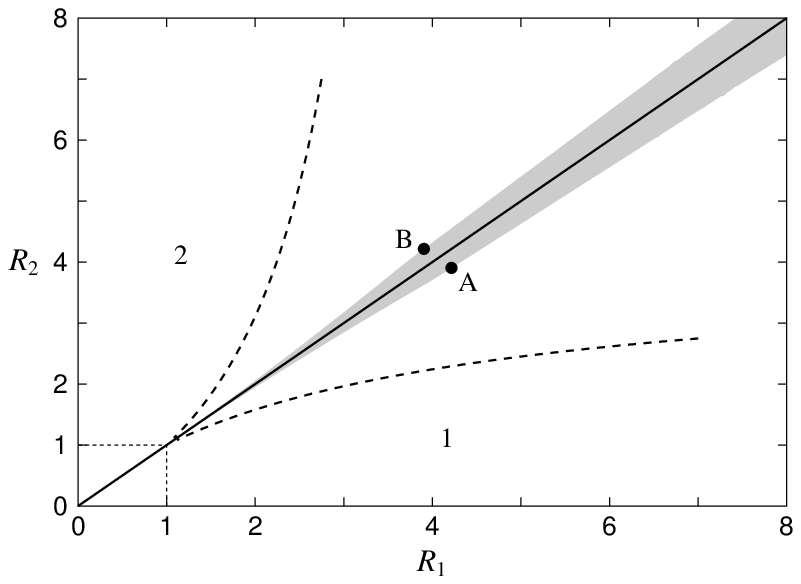}&\includegraphics
{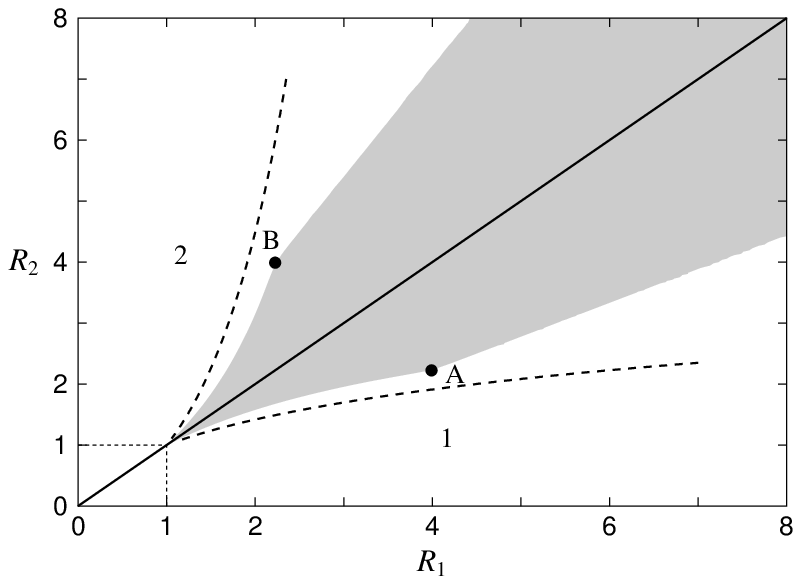}\\
a) & b)\\
\includegraphics
{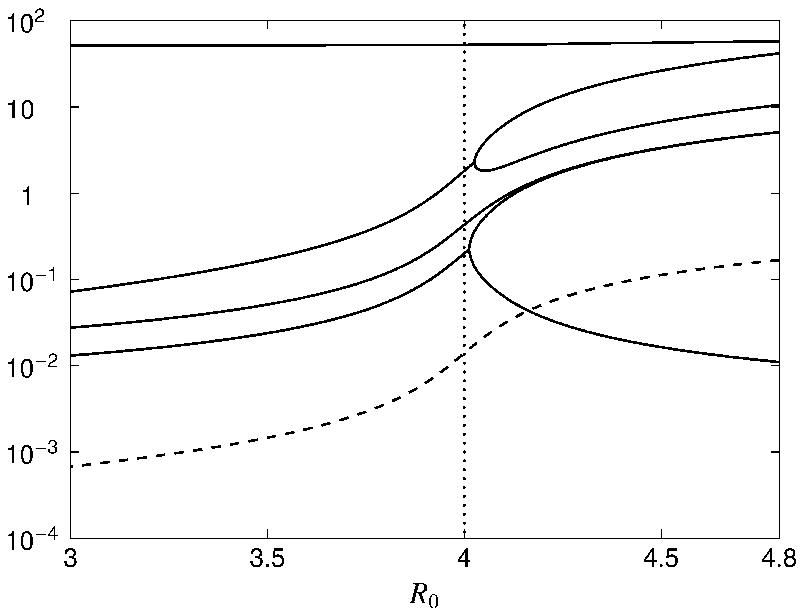}&\includegraphics
{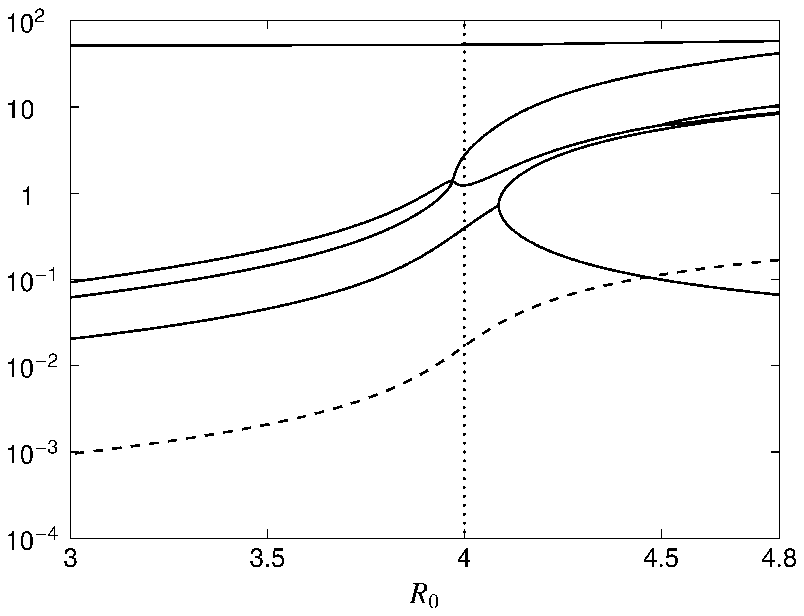}\\
c) & d) 
\end{tabular}
\end{center}
\caption{}
\label{fig4}
\end{figure}

\newpage

\begin{figure}[tbp]
\begin{center}
\begin{tabular}{ccc}
\includegraphics
{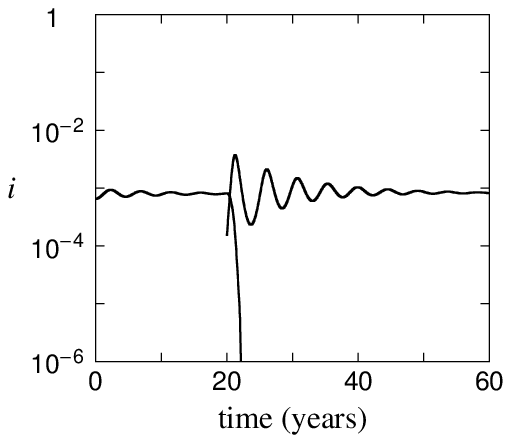}&\includegraphics
{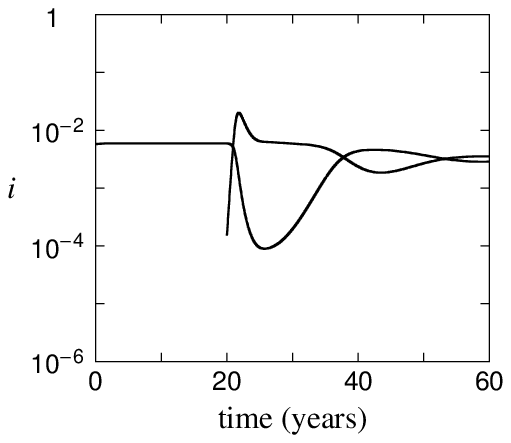}&\includegraphics
{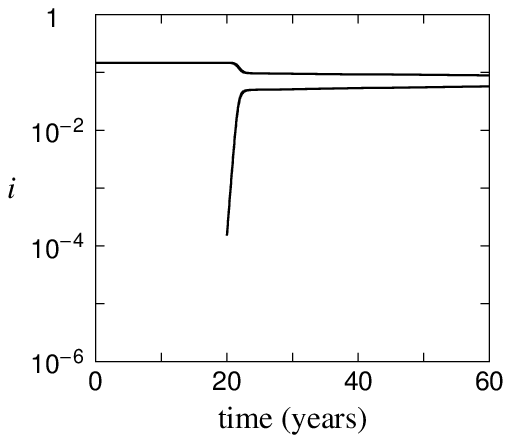}\\
a) & b) & c) \\                                                                                             
\includegraphics
{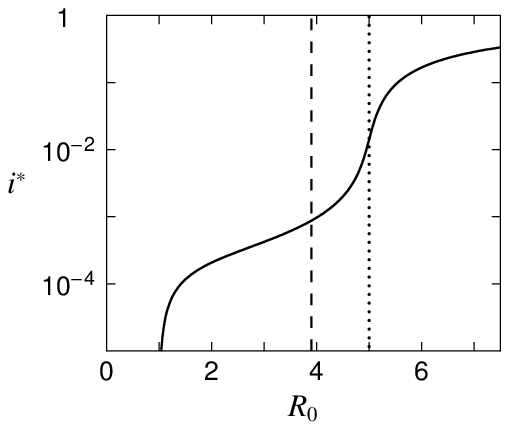}&\includegraphics
{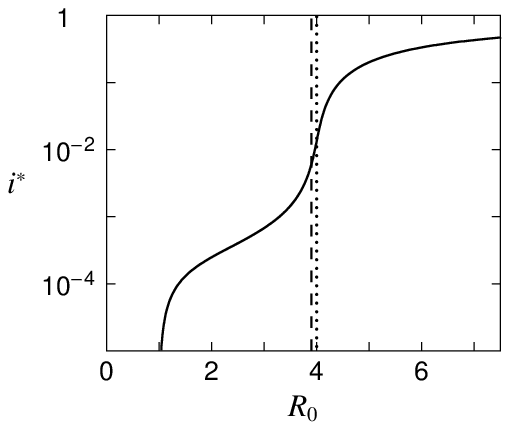}&\includegraphics
{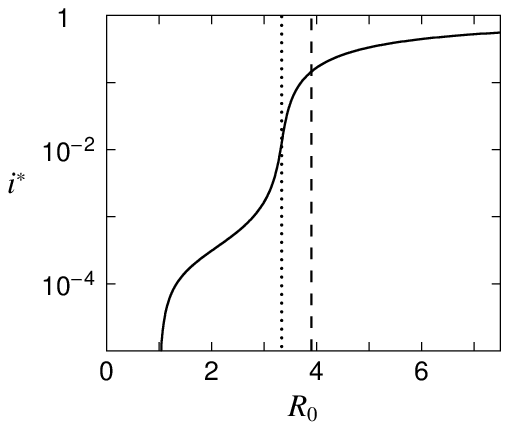}\\                                                                                             
d) & e) & f)
\end{tabular}
\end{center}
\caption{}
\label{fig2}
\end{figure}

\newpage

\begin{figure}[tbp]
\begin{center}
\begin{tabular}{ccc}
\includegraphics
{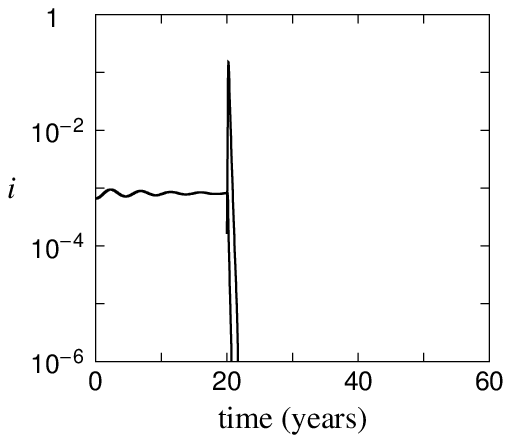}&\includegraphics
{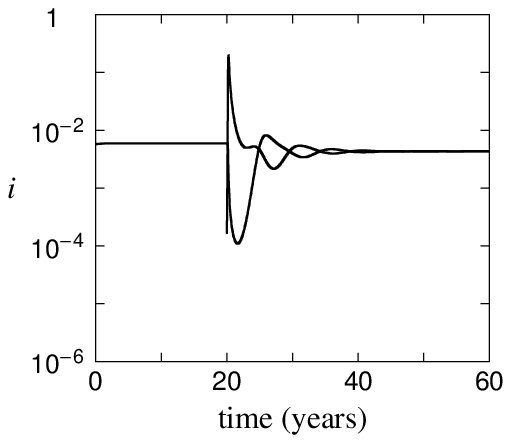}&\includegraphics
{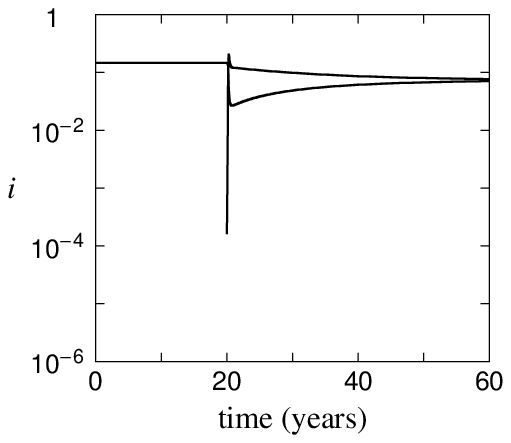}\\
a) & b) & c)\\                                                                                             
\includegraphics
{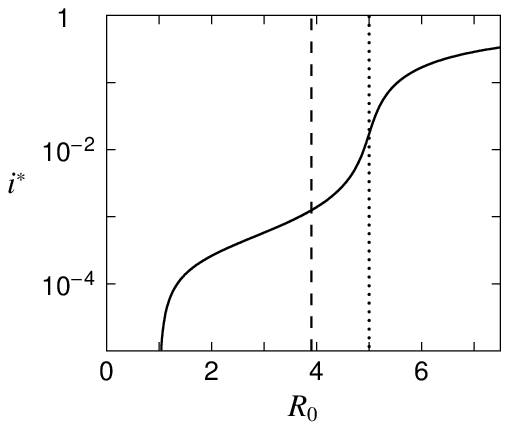}&\includegraphics
{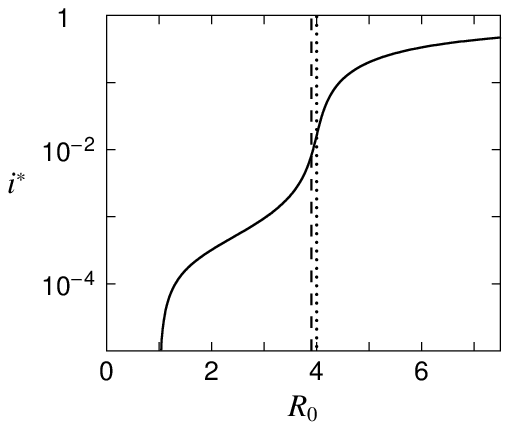}&\includegraphics
{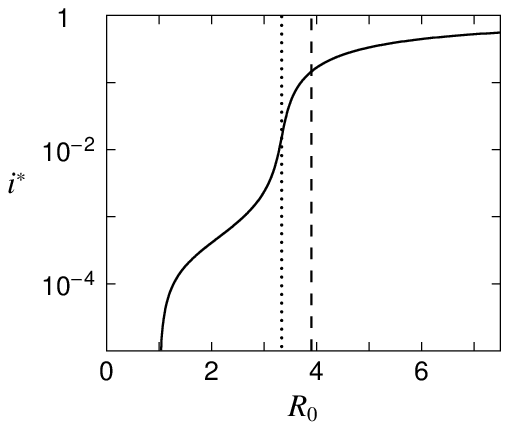}\\
d) & e) & f)
\end{tabular}
\end{center}
\caption{}
\label{fig3}
\end{figure}

\newpage

\begin{figure}[tbp]
\begin{center}
\begin{tabular}{c}
\includegraphics
{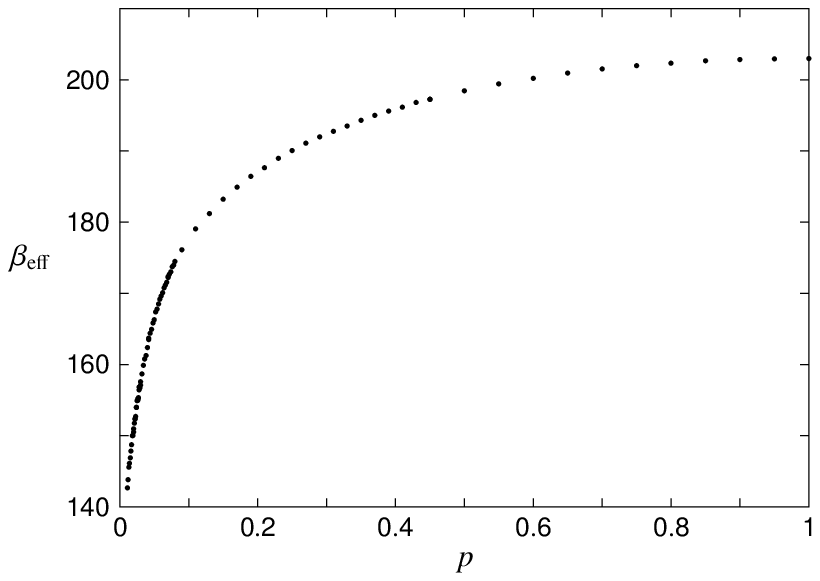}\\
\end{tabular}
\end{center}
\caption{}
\label{fig4}
\end{figure}



\newpage

\begin{figure}[tbp]
\begin{center}
\begin{tabular}{c}
\includegraphics
{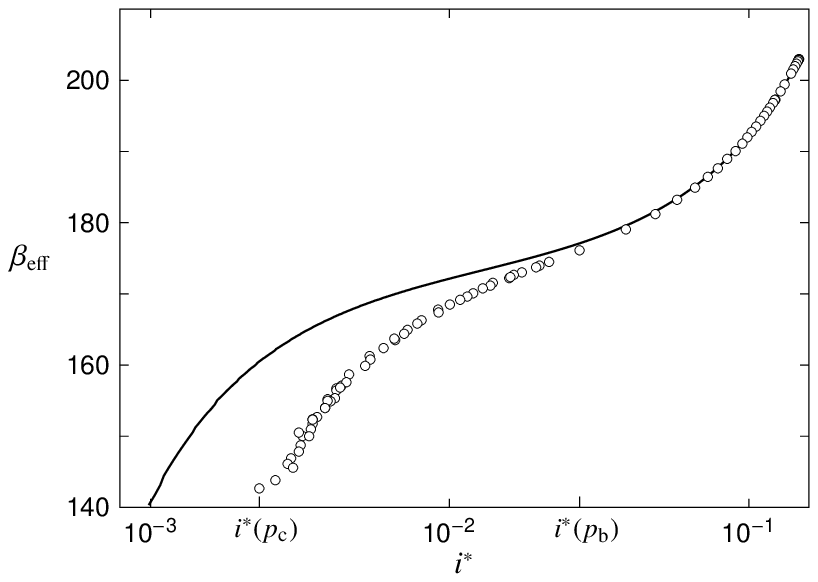}\\
\end{tabular}
\end{center}
\caption{}
\label{fig5}
\end{figure}

\newpage

\begin{figure}[tbp]
\begin{center}
\begin{tabular}{cc}
\includegraphics
{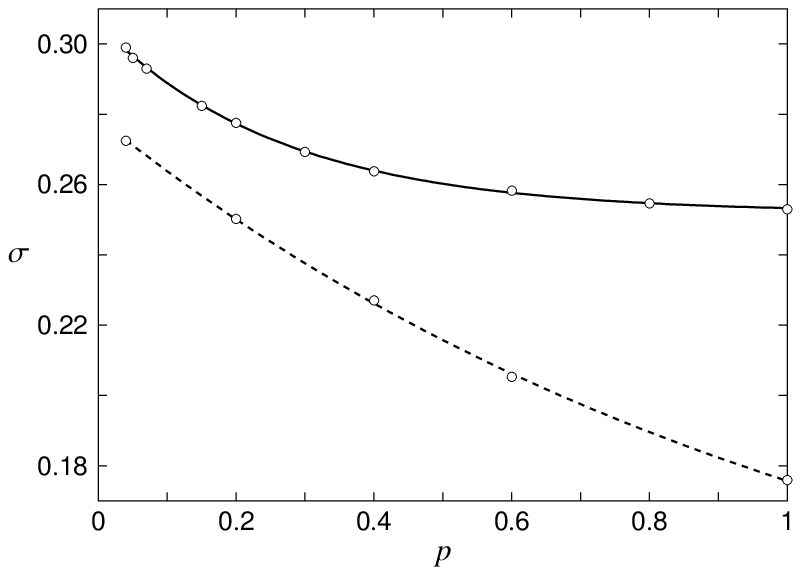}& \includegraphics
{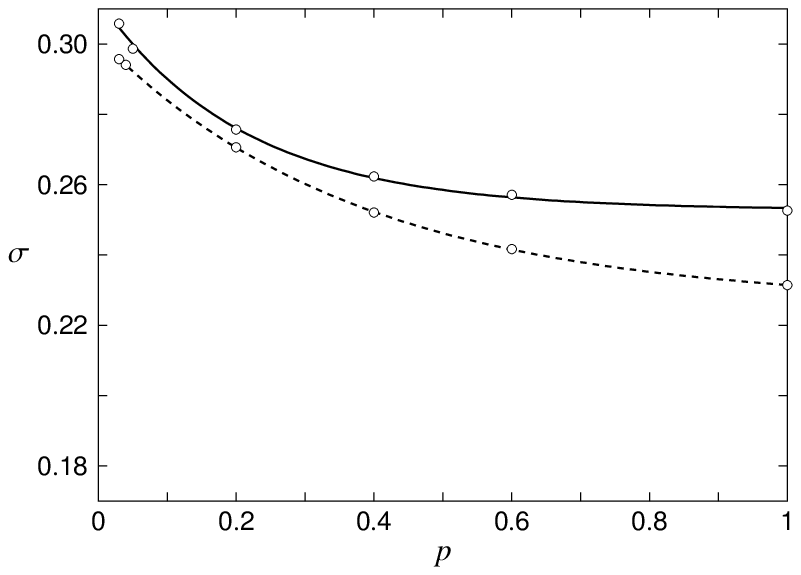}\\
a) & b)
\end{tabular}
\end{center}
\caption{}
\label{fig6}
\end{figure}

\newpage

\begin{figure}[tbp]
\begin{center}
\begin{tabular}{cc}
\includegraphics
{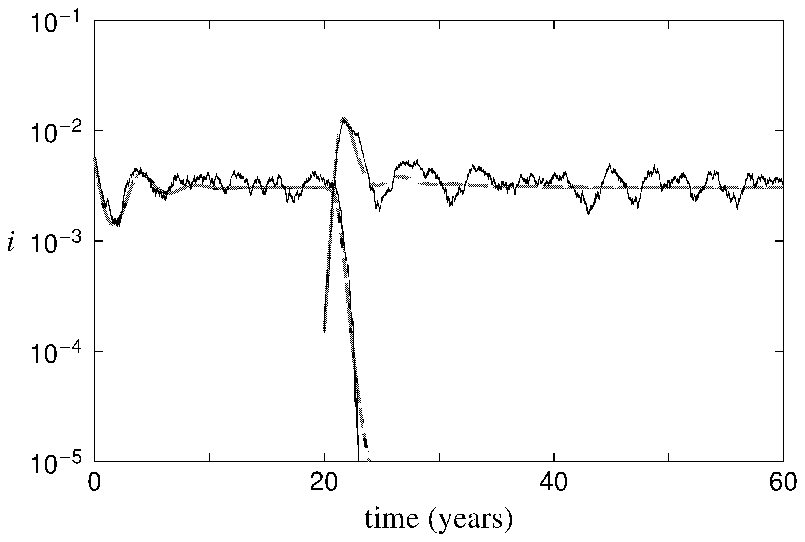} &\includegraphics
{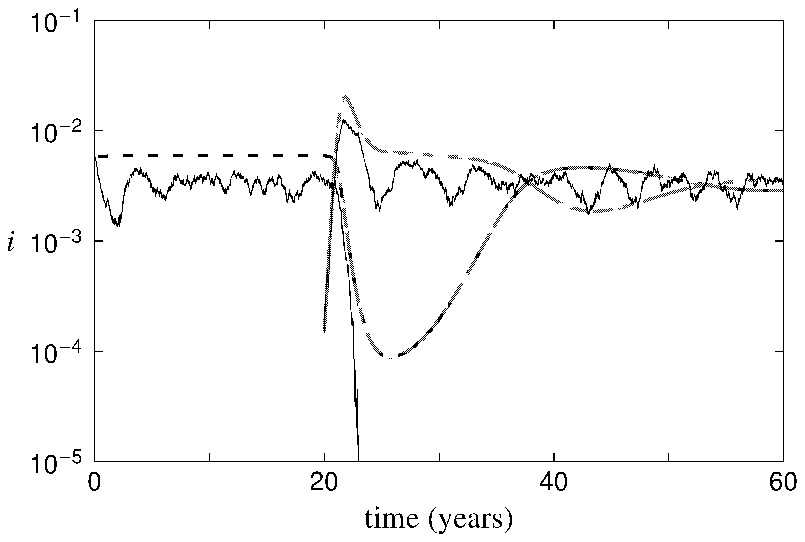} \\
a) & b)
\end{tabular}
\end{center}
\caption{}
\label{fig7}
\end{figure}

\end{document}